# Bell's Theorem: Two Neglected Solutions.

*Louis Vervoort*

*University of Montreal, 29.01.2013 (first version 29.03.2012)*
*louis.vervoort@umontreal.ca, louisvervoort@hotmail.com*

**Abstract**. Bell's theorem admits several interpretations or 'solutions', the standard interpretation being 'indeterminism', a next one 'nonlocality'. In this article two further solutions are investigated, termed here 'superdeterminism' and 'supercorrelation'. The former is especially interesting for philosophical reasons, if only because it is always rejected on the basis of extra-physical arguments. The latter, supercorrelation, will be studied here by investigating model systems that can mimic it, namely spin lattices. It is shown that in these systems the Bell inequality can be violated, even if they are local according to usual definitions. Violation of the Bell inequality is retraced to violation of 'measurement independence'. These results emphasize the importance of studying the premises of the Bell inequality in realistic systems.

## 1. Introduction.

Arguably no physical theorem highlights the peculiarities of quantum mechanics with more clarity than Bell's theorem [1-3]. Bell succeeded in deriving an experimentally testable criterion that would eliminate at least one of a few utterly fundamental hypotheses of physics. Despite the mathematical simplicity of Bell's original article, its interpretation – the meaning of the premises and consequences of the theorem, the 'solutions' left - has given rise to a vast secondary literature. Bell's premises and conclusions can be given various formulations, of which it is not immediately obvious that they are equivalent to the original phrasing; several types of 'Bell theorems' can be proven within different mathematical assumptions. As a consequence, after more than 40 years of research, there is no real consensus on several interpretational questions.

In the present article we will argue that at least two solutions to Bell's theorem have been unduly neglected by the physics and quantum philosophy communities. To make a self-contained discussion, we will start (Section 2) by succinctly reviewing the precise premises on which the Bell inequality (BI) is based. In the case of the deterministic variant of Bell's theorem these premises comprise locality and 'measurement independence' (MI); for the stochastic variant they are MI, 'outcome independence' (OI) and 'parameter independence' [4-6]. Rejecting one of these premises corresponds to a possible solution or interpretation of Bell's theorem - if it is physically sound. In Section 3 we will succinctly review well-known positions, which can be termed 'indeterminism' (the orthodox position) and 'nonlocality' (in Bell's strong sense), and give essential arguments in favor and against them. We believe this is not a luxury, since it seems that quite some confusion exists in the literature: popular slogans such as 'the world is nonlocal', 'local realism is dead', 'the quantum world is indeterministic' are not proven consequences of a physical theory, but metaphysical conjectures among others - or even misnomers. It is therefore useful to clearly distinguish what is proven within a physics theory, and what is metaphysical, i.e. what is not part of physics in the strict sense. All solutions to Bell's theorem are the conjunction of physical and metaphysical arguments.

The first position we will investigate here in more detail (Section 4), and that is usually termed total or 'superdeterminism' is, although known, rarely considered a serious option (notable exceptions exist [6-11]). The negative reception of this interpretation is based on arguments of 'free will' or conspiracy, which are however heavily metaphysically tainted. We will argue that rejection of determinism on the basis of these arguments is in a sense surprising, since it corresponds to a worldview that has been convincingly defended by scholars since centuries; and especially since it is arguably the simplest model that agrees with the facts. (The Appendix gives a condensed overview of the history of this position, where a special place is given to Spinoza.) Its main drawback however – for physicists – is that it seems difficult to convert into a fully physical theory.

In Section 5 we will argue a fourth solution exists, which could be termed 'supercorrelation', and which does not have the latter disadvantage – it is essentially a physical model. In order to investigate its soundness, we will study highly correlated model systems, namely spin lattices. It will be shown that in a Bell-type correlation experiment on such lattices the Bell inequality can be strongly violated; yet these systems are 'local' according to usual definitions [1, 12]. This violation will be retraced to violation of MI. It will be argued that a similar 'supercorrelation' may happen in the real Bell experiment. This will lead us to the conclusion that the premises on which Bell's theorem is based, such as MI, OI and PI, are extremely subtle, and that it is highly desirable to study them in realistic physical systems, not just by abstract reasoning.

Two words of caution are in place. The first is that we will not try to elaborate here a realistic hidden variable theory for quantum mechanics, which seems a daunting task; we are concerned with the much more modest question whether such theories are possible. We are well aware that this implies quite some speculation; but in view of the extraordinary importance of Bell's theorem for physics (and philosophy) such efforts seem justified, especially if arguments can be backed-up by physical models. Second, there exist many highly valuable contributions to the present field, both experimental and theoretical. It would be far outside the scope of this article to review these works;

we could only refer to the texts that were most relevant for the present findings. Let us however start by paying tribute to Bell himself: his texts [1-3] remain landmarks of clarity, simplicity, and precision.

## 2. Assumptions for deriving the Bell Inequality (BI).

For following discussion it will prove useful to distinguish the deterministic and stochastic variant of Bell's theorem. Within a deterministic hidden variable theory (HVT), the outcomes $\sigma_1$ and $\sigma_2$ (say spin) of a Bell-type correlation experiment are supposed to be deterministic functions of some 'hidden variables' (HVs) $\lambda$, i.e.

$$\sigma_1 = \sigma_1(a,\lambda) \text{ and } \sigma_2 = \sigma_2(b,\lambda), \tag{1}$$

where a and b are the left and right analyzer directions. (In the following $\lambda$ may be a set or represent values of fields; the HVs may be split in $\lambda_1$, $\lambda_2$ etc.: all these cases fall under Bell's analysis.) Recall that (1) assumes '*locality*'[1]: $\sigma_1$ does not depend on b, and $\sigma_2$ not on a. In Bell's original 1964 article [1] it is assumed that the mean product $M(a,b) = <\sigma_1.\sigma_2>_{a,b}$ can be written as

$$M(a,b) = <\sigma_1.\sigma_2>_{a,b} = \int \sigma_1(a,\lambda).\sigma_2(b,\lambda).\rho(\lambda).d\lambda. \tag{2}$$

In the most general case however $\rho(\lambda)$ in (2) should be written as a conditional probability density $\rho(\lambda|a,b)$ [4-6]. Indeed, it is essential to realize that from (2) the BI can only be derived if one also supposes that

$$\rho(\lambda|a,b) = \rho(\lambda|a',b') \equiv \rho(\lambda) \text{ for all relevant } \lambda, a, b, a', b' \quad \text{(MI)}, \tag{3}$$

a condition usually termed 'measurement independence' (MI) [5-6,11]. This hypothesis expresses that $\lambda$ is stochastically independent of the variable couple (a,b), for all relevant values of a, b and $\lambda$. There are of course good reasons to suppose that (3) indeed holds. In an experiment with sufficiently rapidly varying analyzer settings [13], creating a spacelike separation between the left and right measurement events, it would seem that the value of $\lambda$ determining $\sigma_1$ on the left cannot depend on the simultaneous value of b on the right (similarly for $\sigma_2$ and a) – at least if one assumes Bell's relativistic locality (see footnote 1). So this argument says that $\lambda$ cannot depend on both a and b, i.e. that MI in (3) holds as a consequence of locality.

---

[1] According to Bell's original [1], a HVT is local iff 1) the force fields the theory invokes are well-localized (they drop off after a certain distance, therefore (1) can be assumed even in an experiment with static settings); and 2) it does not invoke action at-a-distance, i.e. it invokes only influences that propagate at a (sub)luminal speed, in particular between the 'left' and 'right' part of the experiment. Notice this corresponds to an extremely mild locality condition: any known physical system satisfies it.

Before critically analyzing MI in Sections 4 and 5, let us already observe that it has never been rigorously proven that Bell's locality necessarily implies MI. One may well wonder whether this view captures all cases, and whether MI can be violated even in local systems. Shimony, Horne and Clauser [14] observed that 'measurement dependence', i.e. stochastic dependence of λ on (a,b), could in principle arise if the (values of the) variables λ, a, and b have local *common causes* in their overlapping backward light-cones. More generally, it is maybe conceivable that local correlations exist between λ and (a,b) at the moment of measurement which are a remnant of their common causal past; this is a more general variant of the argument in [14] to be discussed in Section 5. The fact is that the counterargument of Shimony et al. against (3) seems to have had little impact in the literature. It has been discussed in a series of articles [14] by Shimony, Horne, Clauser and Bell (reviewed in [15]). All come to the conclusion that (3) should be valid on the basis of a 'free will' argument, a position which seems largely dominant till date. According to the latter view, if λ would depend on a and b, then a and b should depend on λ (due to the standard reciprocity of probabilistic dependence). But the values of a and b can be *freely* chosen by one or even two experimenters; how then could they depend on HVs λ that moreover determine the measurement outcomes $\sigma_1$ and $\sigma_2$ ? Ergo, MI must hold. However, we will prove in Section 5.1. that this 'free will' argument is plainly false.

In sum, assuming (1), the existence of local deterministic HVs, locality, and (3), MI, one derives the Bell-CHSH [16] inequality:

$$X_{BI} = M(a,b) + M(a',b) + M(a,b') - M(a',b') \leq 2 \qquad (4)$$

by using only algebra. Many have summarized that (4) follows from the assumptions of 'HVs and locality' or from 'local HVs'. But, as we showed above, this phrasing is only valid if locality implies MI.

After Bell's seminal work, Clauser and Horne [12], Bell [2] and others extended the original theorem to *stochastic* HVTs. In such a HVT $\sigma_1$ and $\sigma_2$ are probabilistic variables, for which one assumes (instead of (1)) that

$$P(\sigma_1|a,\lambda), P(\sigma_2|b,\lambda) \text{ and } P(\sigma_1,\sigma_2|a,b,\lambda) \text{ exist.} \qquad (5)$$

If $\sigma_1$ and $\sigma_2$ are stochastic variables one has now (instead of (2)) that

$$M(a,b) = <\sigma_1.\sigma_2>_{a,b} = \sum_{\sigma_1=\pm 1} \sum_{\sigma_2=\pm 1} \sigma_1.\sigma_2.P(\sigma_1,\sigma_2|a,b), \qquad (6)$$

where $P(\sigma_1,\sigma_2|a,b)$ is the joint probability that $\sigma_1$ and $\sigma_2$ each have a certain value (+1 or -1) given that the analyzer variables take values a and b (all this in the Bell experiment). Assuming (5), the existence of HVs, and exactly the same condition (3) (MI) as before, it follows from (6) that

$$M(a,b) = <\sigma_1.\sigma_2>_{a,b} = \sum_{\sigma_1=\pm 1} \sum_{\sigma_2=\pm 1} \sigma_1.\sigma_2. \int P(\sigma_1,\sigma_2|a,b,\lambda).\rho(\lambda).d\lambda. \qquad (7)$$

To derive the Bell inequality (4) one has now to make two supplementary assumptions [4-6], usually termed 'outcome independence' (OI) and 'parameter independence' (PI), which are defined as follows:

$$P(\sigma_1|\sigma_2,a,b,\lambda) = P(\sigma_1|a,b,\lambda) \text{ for all } (\lambda,\sigma_1,\sigma_2) \qquad \text{(OI)}, \qquad (8)$$

$$P(\sigma_2|a,b,\lambda) = P(\sigma_2|b,\lambda) \text{ for all } \lambda \text{ and similarly for } \sigma_1 \qquad \text{(PI)}. \qquad (9)$$

Using (8-9) one derives from (7) that

$$M(a,b) = \sum_{\sigma_1=\pm 1} \sum_{\sigma_2=\pm 1} \sigma_1.\sigma_2. \int P(\sigma_1|a,\lambda).P(\sigma_2|b,\lambda).\rho(\lambda).d\lambda, \qquad (10)$$

from which the same BI as before (see Eq. (4)) follows by using only algebra. Note that the original work by Clauser and Horne [12] assumed the so-called 'factorability' condition

$$P(\sigma_1,\sigma_2|a,b,\lambda) = P(\sigma_1|a,\lambda).P(\sigma_2|b,\lambda) \qquad \text{for all } (\lambda,\sigma_1,\sigma_2), \qquad (11)$$

which is however simply the conjunction of (8) and (9). Clauser and Horne justified their assumption Eq. (11) by stating that it is 'reasonable locality condition'; Einstein locality manifests itself in (11) by the fact that $P(\sigma_1|a,\lambda)$ does not depend on b; similarly for $P(\sigma_2|b,\lambda)$ [12, 2]. Since then the factorability condition (11) seems to have become in the literature the *definition* of locality in stochastic systems. (However, even if (11) or OI and PI may be found 'reasonable' in a Bell experiment with spacelike separation between the left and right measurements, one may doubt their general validity – e.g. for the same reasons for which MI may be questioned. We will come back to this point in Section 5.)

Thus, for stochastic HVTs the Bell-CHSH inequality (4) follows from the assumption of MI, OI and PI. Actually, it appears that all known derivations of generalized Bell inequalities are based on assumptions equivalent to (or stronger than) OI, MI and PI [6]. A more generally known phrasing is that the BI (4) follows from the assumption of 'HVs and locality'. But again, here it must be assumed that locality implies MI, OI and PI; an unproven hypothesis.

## 3. The obvious solutions to Bell's theorem: Indeterminism (S1) and Nonlocality (S2).

In the present Section we will have a brief look at well-known positions which may be adopted with respect to Bell's theorem and the experimental results (we do not claim to review all admissible solutions). This exercise has been done before (see e.g. [2,3,8,10,17]), but it still seems useful to highlight some pitfalls; simply recognizing that *all* solutions to Bell's theorem have both a physical and a metaphysical component will already prove helpful. Let us first summarize the discussion of Section 2 in its most precise and presumably least controversial manner. In the case of deterministic HVTs, the BI (4) follows from the following assumptions or conditions (C1-C3):

*The existence of deterministic HVs* (see Eq. (1))  (C1)

*Locality* (see footnote 1)  (C2)

*Measurement independence (MI)* (see (3)).  (C3)

In the case of stochastic HVTs, the BI (4) is derived based on following hypotheses:

*The existence of stochastic HVs* (see Eq. (5))  (C4)

*Measurement independence (MI)* (see (3))  (C3)

*Outcome Independence (OI)* (see (8))  (C5)

*Parameter Independence (PI)* (see (9)).  (C6)

Since in the Bell experiment the BI (4) is violated (e.g. if the particle pair is in the singlet state), and quantum mechanics vindicated, one simply infers that at least one of the assumptions (C1 – C3) must be false; and that at least one of the conditions (C3 – C6) must be false. Rejection of one particular of these assumptions corresponds to one of the admissible interpretations or solutions of Bell's theorem, if it is physically meaningful. If one can legitimately assume that MI, OI, and PI follow from locality then one can resume Bell's theorem in the following condensed way[2]:

*Local HVTs (deterministic and stochastic) are impossible.*  (B1)

This is indeed the phrasing that could resume the work of Bell [1-3] and many others since. Since (B1) is so popular we will first have a closer look at it, but it is important to remember that the conditions (C1-C3) and (C3-C6) are a more precise (and more recent) starting point to analyze Bell's theorem [4-6]; they are at the basis of further solutions. Also recall that (B1) does not only apply to Bell's original spin correlation experiment, but to any entangled state of two quantum systems.

In particular, (B1) says that nature, or in any case a broad class of correlated quantum phenomena, cannot be described by a theory that is *both* local *and* more complete than quantum mechanics. Several articles helpful to understand the full scope of (B1) have derived Bell's theorem without *explicitly* invoking hidden variables [18-20]. Instead of hidden variables they assume that the physical properties $\sigma_1$ and $\sigma_2$ measured in Bell's experiment *have an objective value even before the*

---

[2] (B1) takes the (almost ideal) experimental results into account. Here we do not consider certain so-called loopholes linked to the fact that real Bell experiments would not be 100% faithful tests of the theorem (these loopholes seem to become more and more unlikely).

*measurement:* a hypothesis generally termed 'realism'. These works have thus led to following popular variant of (B1):

*Local realistic theories are impossible.* (B2)

Since the assumptions leading to the conclusions (B1) and (B2) lead to exactly the same type of mathematical inequality (4), it is only logical to suppose that also (B1) and (B2) are equivalent. This is indeed the case, at least for deterministic HVTs[3]. But since (B1) also includes stochastic HVTs, it is more general and more precise than (B2); (B1) would thus explain what 'realistic' in (B2) really means. Moreover, it should be noted that the term 'realistic' as used by the community of quantum philosophers and physicists might give rise to confusion. In the original meaning of the word, as used by the broad community of philosophers, 'realism' is the hypothesis that *the physical world (e.g. physical properties and their values) exists independently of the human mind* [21-22]. Bohr was doubtlessly a realist in the latter, original sense [23]; but he was as surely a non-realist in the sense used in (B2). According to Bohr, the measurement *apparatus* determines the values of quantum properties $\sigma_i$, not the human mind [24]. For these reasons we will rely in this Section on Bell's original phrasing (B1), excelling in clarity, rather than on (B2).

Some authors have concluded from the observation that quantum mechanics rules out 'any' local HVT (understood: both deterministic and probabilistic), that the conflict arises because of the locality condition alone: it would then be proven by Bell's theorem (and the experiments) that quantum systems are necessarily nonlocal[4]. This interpretation of Bell's theorem is widespread (see e.g. [25] and [8] Chap. 6). However, it is crucial to realize that it is not a proven result: it is one possible interpretation (see S2 below), and certainly not the only admissible position (unless one redefines 'nonlocal' of course). Moreover, it appears that the nonlocality that is invoked by these authors is of a strange kind: it cannot be used for superluminal signal exchange [8]. Alain Aspect and Asher Peres, for instance, term the photon pair of the Bell experiment a 'single non-separable' or a 'single indivisible, nonlocal' object ([25], [8] Chap. 6). But does this really help to understand the correlations in the Bell experiment ? In Bell's work nonlocality straightforwardly refers to superluminal interactions or signals; but not so in the latter interpretation of nonlocality. It seems that Peres attempts, in certain texts (e.g. [8] Chap. 6), to explain quantum nonlocality by the intuitive idea that in 'one indivisible object', if one part 'feels' something, any distant part of the same object immediately also 'feels' something. But this is not a description of what happens in any normal solid object, in which an influence (a force) exerted on one part propagates to any other part at subluminal speed.

---

[3] This may be shown as follows. In a deterministic HVT the physical properties $\sigma_1$ and $\sigma_2$ have a value before each instance of measurement (since $\sigma_i$ is determined by, i.e. a function of, $\lambda$); conversely, in a realistic theory $\sigma_1$ and $\sigma_2$ have a value even before measurement and one can, implicitly or explicitly, index that value by a hidden variable or index.

[4] Here 'nonlocal' cannot mean 'entangled' because entanglement is of course not proven by Bell's theorem. In information-theoretic texts 'nonlocal' is often synonymous to 'violating the BI', but in our discussion the term is obviously not used in that way.

Obvious interpretations of Bell's theorem, then, are an immediate consequence of (B1).

*S1. The Standard Solution ('indeterminism')*. The orthodox position, probably adopted by a majority of physicists, is to reject Bell's 'hidden variable' hypothesis (C1 and C4). So according to this position *there are no hidden variables completing quantum mechanics* (for the Bell experiment and other entangled systems), *not even in principle*. As argued above, rejection of deterministic HVs amounts to rejecting a special kind of 'realism', more precisely to rejecting the thesis *that the values of the quantum properties $\sigma_1$ and $\sigma_2$ (of Bell's experiment) exist even before their measurement*. This is the orthodox position because it has in essence been anticipated by Bohr decades before Bell's discovery [1] – it is part of the original Copenhagen interpretation ! [23-24, 8] Indeed, in 1935 Bohr had dismissed the EPR paradox by invoking this position. As we read it, Bohr's anti-EPR argument [24] can be summarized as follows: measurement brings observables into being through an inevitable interaction with an observing system; if two observables cannot be measured simultaneously, they do not exist simultaneously. Bohr might have similarly argued that the values of the quantum properties $\sigma_1$ and $\sigma_2$ of Bell's experiment do not exist before measurement, are not determined, i.e. Bell's theorem (B1) or (B2) is not valid.

(It seems that there is one small cloud that stains this perfect picture, namely the fact that this argument à la Bohr does not immediately explain why *stochastic* HVTs would be inconceivable. But advocates of S1 could invoke a remarkable result by Fine [26], proving the equivalence between the existence of stochastic and of deterministic HVTs. And we will see further that Bohr might have resorted to a second argument, linked to the 'contextuality' of the Bell experiment, i.e. the importance of considering the whole experimental set-up including the *two* analyzer settings a,b: see e.g. footnote 7.)

The main arguments in favor of the Standard Position are, it seems to us, the following. First, it saves locality in Bell's sense. And of course, it is part of the Copenhagen interpretation which has proven itself countless times since its conception. Here one may however observe that the Copenhagen interpretation contains theses that have mainly a physical content (such as Born's rule) which indeed are admirably confirmed by experiment, but also metaphysical theses – and rejecting HVs obviously belongs to these extra-physical hypotheses. Rejecting any yet to discover HVs reminds of a slogan as "we talk about what we can measure or calculate; the rest does not exist" – a slogan which summarizes an axiom of the positivist philosophy, by which Bohr may well have been influenced [27, 2]. In sum, in a sense position S1 espouses well the daily practice of mainstream physics; *in a sense* it seems to make extra-physical commitments that are minimal. However, it is not part of physics in the strict sense. Proponents of other positions may very well continue to inquire about the perfect correlations of the EPR/Bell experiment. Probability theory does not prohibit that probabilities (which these correlations are) are considered as resulting from underlying causal mechanisms. The examples of physical systems in which probabilistic behaviour can be retraced to deterministic laws, are countless. And in a different sense, S1 and the Copenhagen interpretation become quite spectacular, and metaphysically heavily loaded. Indeed, S1 can be restated as follows: when, at the moment of measurement of a quantum property $\sigma$ like spin (in the Bell experiment) one obtains a certain value (+1 or -1), this value is the result of *absolute hazard*, in the sense that it will never be possible, not even in principle, to explain this result. No theory can ever be constructed going beyond the statistical predictions of quantum mechanics (only $P(\sigma)$ exists). In other words: we are in the presence of events that have no cause; the microscopic world is full of such 'indeterministic' events; and quantum mechanics is the 'final' theory for such events (in the sense stated). (Note that what S1 claims for the properties measured in a Bell experiment, the Copenhagen interpretation claims for *all* quantum properties, except for those that are in an eigenstate of the measured property.)

*S2. The Non-Standard Solution ('nonlocality' in Bell's strong sense).* The non-standard approach is to conclude from (B1) that locality (in Bell's sense) is violated in nature, i.e. that superluminal influences (forces) exist. Needless to say, this is a speculative, unorthodox solution since it violates relativity theory. However, it is possible in principle ([2], [8] p. 171). A well-known example of a non-local HVT is Bohm's theory [28]. Note that it seems that of all the solutions reviewed here S2 and S4 below are the only ones that could in principle be proven. Also, the superluminal force field needed for S2 could be ultraweak and dynamically enhanced by nonlinear dynamics, as shown in [29]. An ultraweak force field may have, till date, escaped from detection.

It will be no surprise that for many people both S1 and S2 remain unsatisfactory, for the reasons stated. In the next Sections we will investigate two further positions, which aim at avoiding the 'unpleasant' features of S1 and S2.

**4. First Neglected Solution: Superdeterminism (S3).**

That 'total determinism' or 'superdeterminism' (S3) offers a solution to the Bell impasse has been observed by a few physicists, soon after [1], starting by Bell himself [2, 14]. Besides in interesting analyses by authors as Brans [7], Peres [8], 'tHooft [9], Khrennikov [10] and recently Hall [6,11], S3 has until now been considered a completely implausible solution. However, the debate may be unduly unbalanced, as we will argue now. Whereas S1 and S2 accept Bell's original no-go theorem (B1), total determinism (S3) questions its derivation; or rather, starts from the more precise analysis (C1-C3), and recognizes that MI (C3) is an essential assumption of any derivation of the BI.

In short, total determinism assumes 1) that *any* event, property or variable is determined in the sense (1), *including the choices 'a' and 'b' of the analyzer settings* in a Bell experiment; and 2) that there is, when going back in time, a contraction of the 'causal tree' (the collection of all events) to a small space-time region – which is nothing more than the hypothesis of the Big-Bang. If this view is correct, it immediately follows that $\sigma_1$ and $\sigma_2$ must be determined by some cause $\lambda$, but also that a, b and $\lambda$ should have *common causes* (since the 'world cone' converges to say a point). Now, as we saw in Section 2, in that case $\lambda$ may stochastically depend on (a,b) and MI in (3) does not necessarily hold, even in a fully local world. This point has first been made by Shimony et al. [14], it seems. If MI (C3) does not hold, Bell's no-go theorem cannot be derived, i.e. there may be local HVTs reproducing the quantum statistics.

Thus, solution S3 considers the world as 'superdeterministic', in that even our choices of parameter settings are determined and *in principle* linked to virtually all other physical properties through a retracting world cone. Because this position seems in contradiction with a classic conception of 'free will' [15], Bell termed S3 a 'mind-boggling' or 'conspiratorial' option [2]; David Mermin speaks of 'the most paranoid of conspiracy theories' [19]. These negative verdicts seem to have had a lasting influence on the community of quantum physicists and philosophers. But since these arguments are not strictly part of physics, but of philosophy, it seems inappropriate to disconnect the discussion from the most relevant philosophical theories on the matter (in the remainder of this Section and in the Appendix we try to provide at least a mini-introduction, be it extremely condensed). In particular, it deserves to be emphasized that above verdicts, presented as patently commonsensical, neglect a worldview that is cogent, well-documented and widespread[5] outside the quantum community. Determinism is at the heart of philosophical debate since millennia. One of the philosophers who in our personal opinion defended determinism best is Spinoza, who put it at the basis of his system [30-31]. According to Spinoza (and a very considerable part of all philosophers having studied the question) determinism is not antagonistic to free will, just to a simple conception of it.

---

[5] Total determinism seems to be a popular philosophy. For what it is worth, here is the result of a little survey we did, one among about 20 physicists, experts of the foundations of quantum mechanics, one among about 20 philosophy students. In both cases, about 40% of participants said to be in favour of determinism, 60% in favour of indeterminism. (In a third group (30 p.), after a defense of determinism, the ratio was rather inversed.) These surveys were casual and have of course no pretension.

Let us go once more over the arguments. On the one hand it is just normal that Bell [1] started from (implicitly) assuming measurement independence, for at least two reasons. First, it seems that the practice of physics is only possible because separated subsystems can be described by physical parameters (such as λ, a, b) that belong *only* to particular subsystems, and not to all systems; at any rate, 'systemic' descriptions are customary in physics, if not the only possible ones. More importantly, in the Bell experiment the analyzer positions a and b can be chosen and set by an experimenter. How then could they be determined by other variables - variables that moreover would also fix properties of the particle pairs ? That seems too much of a violation of free will - at least, such is the dominant position in the Bell literature. However, from another point of view, assuming MI is not innocent - actually, it arguably corresponds to an uneconomic worldview. Indeed, recall that one of the initial motivations of Bell, just as of Einstein, was to investigate whether quantum mechanics could be made deterministic, as classical theories. Now start, just as Bell, from the hypothesis that certain quantum properties σ, intervening in the Bell experiment, are deterministic (determined, caused, by yet hidden variables). *If one wants to reason within the simplest, most economic model*, then one should also assume that *all other* physical properties are determined (a worldview with both deterministic and indeterministic quantities needs two categories). Still within this most simple worldview or ontology, human beings are also determined physical beings: their actions can be described – in principle, not in practice – by deterministic properties. In short: *all* events (or systems), whether of animate or inanimate origin, are caused by previous events; which are caused in turn by still earlier events, etc.. Now, if one takes the Big Bang theory into account, it would appear that the idea of a universal causal 'branching' or 'network' between events, originating at the Big Bang and connecting (almost) 'anything to anything', is a quite natural conclusion. In sum, measurement independence appears to be in contradiction with a simple ontology, the mentioned 'total' or 'superdeterminism' (actually, determinism would suffice as a term). As is well-known, free will *and* probabilities are explained in this model as 'emerging' due to our unavoidable lack of knowledge of all causal factors. On this view free will is a perception, imparted by our obvious *feeling* to be free – a feeling that surely is immensely functional but still might be an illusion.

So, what seems at first glance, from our daily point of view of free agents, a paranoid conspiracy theory, becomes from another point of view a quite reasonable hypothesis. As recalled in the Appendix, this point of view has been advocated since centuries, and probably millennia, by countless scholars. In this context, it seems that terming S3 'conspiratorial' is rash. Authors who call

S3 'conspiratorial' do so because they look at this position from following angle: an operator (or an automat) doing the Bell experiment would be determined to set, in function of each particle pair, the polarizers in just these positions so as to make the result coincide with the quantum prediction. But this seems an anthropocentric point of view. Determinism basically only says that *every* event happens according to deterministic laws – full stop. Then, if we do an experiment and obtain a result obeying quantum laws these events all *must* be determined *and* linked (see above); in other words quantum mechanics is an effective theory (as many others) having at least in principle a subjacent explanation. This seems a direct consequence of logic and the above simple hypotheses, not of conspiracy[6]. As so often, conspiracy is in the eye of the beholder. All depends from which assumptions one starts.

In conclusion, a third position (S3) w.r.t. Bell's theorem is conceivable, namely, in short, the assumption of total determinism, implying a (local) violation of measurement independence (C3). This position saves locality, does not violate any known physical law, and leaves open the possibility to complete quantum mechanics in the EPR/Bell sense, as in Eq. (1) or (5). It points however to a *very different kind of non-locality*, namely a universal connectedness of virtually all systems (including human beings), due to a receding world cone. Of course, if S3 would be true, the truly mind-boggling thing would be that quantum mechanics and Bell's theorem allow us to discover this ancient and universal link between virtually all objects; and to corroborate a millennia old philosophy. We believe that the main argument in favor of superdeterminism is that it corresponds to the simplest worldview, based on the fewest concepts (and should one not adopt the simplest theory agreeing with the facts ?). Its main drawback is that it may be difficult to directly transpose it in a physical theory. (At least this is Bell's position, see his last article in [14]. The reason invoked is that theories that describe both our choices *and* Bell experiments by explicitly exhibiting common parameters are doubtlessly impossible. Other people are not impressed [9].)

We believe however that a solution exists (S4 below) that is essentially physical, i.e. that may be more easily backed-up by a new physical theory.

---

[6] A well-known theory on conspiracy theories [32] proposes following ground for people believing in conspiracy. In short: extraordinary effects call for extraordinary causes. In the face of events or 'coincidences' that are *perceived* as formidable, people would have a tendency to look for formidable explanations: a conspiracy by higher powers (or simply the powerful). Now, exactly this theory [32] might apply to people calling S3 a 'conspiracy theory' (!): they perceive S3 as too formidable to be true, and believe that only higher powers – a conspiracy – can explain what S3 proposes. In this context, see also Spinoza [30], who denounced fallacious reasoning of a quite similar type. He analyzed in particular the case of his contemporary fellows, who, in view of the perfection of the world and the quasi-infinite potential of harmonious interaction it offers, concluded that it surely must have been made for them by a higher power. But according to Spinoza's determinism, *and modern biology*, nature and mankind evolved in a lawful way so as to *necessarily* be in some kind of harmony – no divine plan is needed. In sum: anthropocentric reasoning is widespread and often wrong.

## 5. Second Neglected Solution: Supercorrelation (S4).

### 5.1. Measurement dependence through past interaction (S4a).

For convenience we will collect *two* potential solutions under the term 'supercorrelation': it will be seen they both invoke stronger (or other) correlations between the system variables $\sigma_1$, $\sigma_2$, $\lambda$, a, b than the other positions do. More precisely, we propose here solutions that refute MI in (C3) *but not through superdeterminism*; and OI in (C5).

Needless to say, abstract reflection on HVTs is perilous – we know by definition almost nothing about them, and it is difficult not be guided by some preconceptions; many people will have very different ideas about what they may look like. In the present Section I am guided by some early attempts to construct realistic HVTs for (certain aspects of) quantum mechanics, in which the HVs are (values of) *fields*. (As far as I know these are at present the most promising candidates.) For instance, in [33-35] the essential HV is a stochastic zero-point field that imparts Brownian motion to quantum particles, from which on average the standard quantum statistics would emerge. Ref. [34] mentions, as one of its sources of inspiration, recent and spectacular experiments by Couder et al. [36], in which quantum behavior (e.g. double slit interference) is reproduced by macroscopic particles, namely oil droplets. The latter are excited by an external field (the vibration of an oil bed) imparting Brownian motion to the droplets. There seems to be a common denominator in these theories [33-35] and experiments [36], a kind of 'contextuality', namely the fact that the precise shape of the (zero-point) field ($\lambda$ for us) depends on the 'context', i.e. the boundary conditions of the *whole* experimental set-up including the parameters of *all*, even remote, detectors. For instance, the experiments [36] impressively show that the wave field guiding a particular oil droplet through slit 1 is determined by the geometry of *both* slits; a feature also present in the HVT for the quantum version of double slit interference of Ref. [34]. If a similar contextual $\lambda$-field would exist in the Bell experiment, then MI in (3) would *not* be satisfied in general – the field $\lambda$ may well depend on (a,b) – at least if the detector settings are static. Note that this violation of MI (at the moment of measurement) would come about not through common causes between $\lambda$ and (a,b), i.e. through superdeterminism, but through an earlier interaction between the field $\lambda$ and the analyzers (a and b). Let us emphasize that the experiments of Couder et al. show that one would be ill-advised to consider such a contextuality as far-fetched: it has now been proven to exist even in macroscopic systems, *and* to lead to quantum-like behavior (including tunneling and quantization of angular momentum) [36].

What if in a Bell experiment a spacelike separation between the left and right measurement events is imposed, as is the case in the most sophisticated experiments [13, 25] ? Then MI seems harder to overcome, but not as untouchable as often believed. First of all, note that these experiments [13, 25] do not really use changing analyzer directions (e.g. rotating analyzers), as Bell repeatedly advocated, starting in his first paper [1]. They achieve spacelike separation between measurement events by using random switches and sufficiently retarding the choice between two *static* analyzer directions (a and a' on the left, b and b' on the right). But then it seems conceivable, within the kind of contextual theories just mentioned, that a hidden field accompanying or describing the particles exhibits four modes each depending on a left and right analyzer direction, so labeled by (a,b), (a',b), (a, b') and (a', b'); and that when the random switches of the experiment chose e.g. (a,b) they select the λ-mode depending on (a,b) (similarly for the 3 other modes) – a resonance phenomenon. In other words, it seems that in this experiment the λ-field may depend on (a,b), i.e. that MI in (3) may be violated, notwithstanding the spacelike separation. As we will see further this argument applies a fortiori to OI. On the above view, it is essential that experiments be done with one polarizer on both sides, each changing its polarization direction rapidly enough (ideally by rotation). This may however be difficult to realize.

Before looking at a model system, I believe it is important to emphasize that 'measurement dependence (through past interactions)' (or 'contextuality') is fully compatible with a precise application of probability theory. As is well known, the interpretation of probability is a surprisingly subtle topic debated among all fathers of probability theory, such as Laplace, Kolmogorov, von Mises etc., and countless scholars since [10]. In recent studies, following von Mises, it has been stressed that 1) probabilities belong to experiments and not to objects or events per se, and 2) that any probability depends *at least in principle* on the 'context' including *all* detector settings of the probabilistic experiment [10, 37]. (This is a position that Bohr held regarding quantum systems, but that arguably holds also for classical probabilistic systems [37]). According to this analysis, then, ρ(λ) in (2) and (7) should in principle be considered as ρ(λ|a,b). This seems the essential idea of several papers criticizing Bell's theorem (see e.g. [10, 38-39] and references therein)[7]. In the present paper we focus on physical arguments on how such a measurement dependence can come about.

Indeed, it seems essential to demonstrate the above ideas in existing physical systems. Theoretical work on the justification or rejection of MI, OI and PI has until now been restricted to mathematical considerations and information-theoretic toy models (for a recent review, see [6]). As far as we know no realistic physical systems have been investigated. The initial question of this Section was: could there be a local HVT for the Bell experiment that violates MI but that *does not*

---

[7] It seems noteworthy that this is an argument that even Bohr might have liked: he often stressed that a quantum system includes the whole measurement equipment, due to the complementary nature of such arrangements.

*invoke common causal factors for λ, a, b*, so that doesn't rely on superdeterminism ? Although we cannot provide a full-blown (and not ad-hoc) HVT, we can exhibit a physical system, well-known from classical statistical mechanics, that violates the BI, that is local, and in which MI is violated without superdeterminism, i.e. even if we assume free will in the usual sense. That local (and classical) HV systems can strongly violate the BI, is a surprise in itself.

We are looking for model systems that are strongly correlated, since for those it is a priori clear that MI, OI and PI (Eq. (3), (8-9)) may not necessarily hold. We want to perform a Bell-type correlation experiment on such a system and verify whether the BI holds. One of the most studied models in statistical mechanics is the Ising model (originally proposed to investigate ferromagnetic phase transitions in low-dimensional electron systems [40-41]). Its Hamiltonian is:

$$H(\theta) = - \sum_{i,j} J_{ij} \cdot \sigma_i \cdot \sigma_j - \sum_i h_i \cdot \sigma_i. \qquad (12)$$

Here the $J_{ij}$ represent the interaction between spin i ($\sigma_i = \pm 1$) and spin j and are responsible for 'cooperative' behaviour and long-range correlations (positive $J_{ij}$ induce lowering of the energy if the spins are aligned). $J_{ij}$ ranges over first neighbours of N particles (electrons, ions, …) sitting on a D-dimensional lattice. The $h_i$ are local magnetic fields and θ is the N-spin configuration {$\sigma_1, \sigma_2,…, \sigma_N$}, occurring according to the usual Boltzmann distribution. Note that this is really a classical system (in quantum versions of (12) the $\sigma_i$ are Pauli matrices, see (17)) but that in some anisotropic magnetic materials, when only one spin direction (z) matters, (12) coincides with the quantum description. Also note that the 'interaction' between the spins is of course not a direct spin-spin interaction; the first term in (12) arises through a Coulomb potential combined with the Pauli exclusion principle[8]. Finally, the $\sigma_i$ do not even need to be spins (they can represent atomic occupation in a crystal or a lattice gas, deviation from equilibrium position in a network of springs, etc.): the Ising Hamiltonian is ubiquitous in physics.

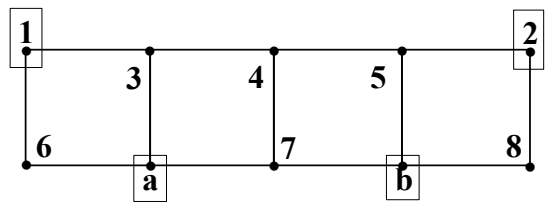

Fig. 1. 10 spins on a square lattice.
Each node contains a spin $\sigma_i$ (i = a,b,1,…,8), which can be up or down.

---

[8] The quantum treatment shows that the $J_{ij}$ correspond to the exchange integrals $\int \psi^*_{ab} \cdot V \cdot \psi_{ba} \cdot d^3x_1 \cdot d^3x_2$, with V the Coulomb potential and $\psi_{ab}(x_1,x_2) = \psi_a(x_1) \cdot \psi_b(x_2)$ (where $\psi_{a(b)}(x_{1(2)})$ are the single electron eigenfunctions located at $x_1$ and $x_2$ respectively) ([41] Chap. 7).

Let us consider a square lattice (Fig. 1) of N = 10 spins (of electrons, ions,…) interacting with first neighbours only. At given temperature $1/\beta$ a configuration $\theta = \{\sigma_a, \sigma_b, \sigma_1,…, \sigma_8\}$ occurs with the Boltzmann probability (or configuration probability) $P(\theta)$ given by:

$$P(\theta) = e^{-\beta H(\theta)} / Z, \text{ with } \beta = 1/kT \text{ and } Z = \Sigma_\theta\, e^{-\beta H(\theta)}, \text{ the partition function.} \quad (13)$$

It is then well-known and straightforward to calculate that this system shows *full pairwise correlation* in the sense that

$$P(\sigma_i=\varepsilon, \sigma_j=\delta) \neq P(\sigma_i=\varepsilon).P(\sigma_j=\delta) \quad \text{for all } i,j \leq N = 10 \text{ and } \varepsilon, \delta = \pm 1. \quad (14)$$

As an example, $P(\sigma_i=+1, \sigma_j=-1)$ is calculated as $\Sigma_\theta P(\theta)$ where the sum runs over the $2^8$ 10-spin configurations $\theta$ in which $\sigma_i=+1$ and $\sigma_j=-1$. Since each term in the sum involves the energy $H(\theta)$ given by (12) the calculation is easily done by numerical simulation. Similarly, in the following any probability $P(\eta)$ with $\eta$ an m-spin configuration (m≤10) is calculated by:

$$P(\eta) = \sum_{\theta(\eta)}^{2^{10-m}} P(\theta), \quad (15)$$

where the sum runs over the $2^{10-m}$ 10-spin configurations $\theta(\eta)$ that 'contain' $\eta$.

We are interested in producing an analog of the Bell experiment in which quantum probabilities as $P(\sigma_1,\sigma_2|a,b)$ are completed or explained by additional variables $\lambda$, by summing over probabilities as in (5) – remember that is what we ask of a stochastic HVT[9]. Consider then an ensemble of 10-spin lattices as in Fig. 1, all at the same temperature, and measure for each lattice the value (±1) of $\sigma_1, \sigma_a, \sigma_2, \sigma_b$ (here $\sigma_a$ and $\sigma_b$ mimic the analyzer variables a and b of the Bell experiment; $\sigma_3, \sigma_4,…, \sigma_8$ can be considered hidden variables)[10]. Measurement on a large ensemble allows to determine the joint probabilities $P(\sigma_1=\varepsilon, \sigma_2=\delta | \sigma_a,\sigma_b)$ ($\varepsilon,\delta = \pm 1$); these probabilities can be determined for any of the 4 possible couples $(a,b) \equiv (\sigma_a,\sigma_b) = (\pm 1,\pm 1)$ by postselecting 4 subsensembles from the total run – exactly as in the real Bell experiments. With these probabilities the average product $M(a,b) = <\sigma_1.\sigma_2>_{a,b}$ for the 4 couples (a,b) can then be determined by using Eq. (6). And finally, putting a≡b≡+1 and a'≡b'≡ −1, the quantity $X_{BI}$ in (4) can be experimentally determined (and calculated) for each parameter set ($J_{ij}$, $h_i$) (we always take $\beta = 1$, a common value [40]).

---

[9] Also, it plays no role whether these $\lambda$ are classical or quantum-like, a case we will investigate elsewhere [42], as explicitly mentioned by Bell (see [2]).
[10] If one wants to push the analogy with the Bell experiment further, suppose that Alice measures $\sigma_1$ and $\sigma_a$, and Bob $\sigma_2$ and $\sigma_b$.

Note that this system is local in any usual sense [1, 12]. It is local in Bell's sense [1] because there is no interaction (and a fortiori no superluminal interaction) between the 'left' and 'right' sides of the lattice, i.e. between ($\sigma_1,\sigma_a$) and ($\sigma_2,\sigma_b$). Indeed, the interactions $J_{ij}$ range only over first neighbors ($J_{ij}$ is taken = 0 otherwise). Importantly, it is also local in the sense that the Clauser-Horne factorability condition (11) is satisfied, with $\lambda = (\sigma_3,\sigma_4,...,\sigma_8)$ or any subset thereof, as is easily calculated by using (15). Recall that (11) serves as the usual definition of locality in stochastic systems. Further, it is easy to calculate that in a lattice in which there *is* an interaction $J_{ij} \neq 0$ between some 'left' or 'right' spins (e.g. between $\sigma_1$ and $\sigma_b$), also the locality condition (11) fails to hold [42].

It is then surprising, to some point, that in this Bell-type experiment on a local (and classical) system the BI (4) can be strongly violated. Violation occurs for wide ranges of system parameters ($h_i$, $J_{ij}$) (as also for a variety of other 1D and 2D geometries we investigated). Indeed, by numerical simulation one finds e.g. for the values $h_i$=1 (all i), and $J_{ij}$=1.4 (first neighbours), that $X_{BI}$ = 2.24 > 2 (the value 2.24 is at least a local maximum). If one allows the magnetic fields $h_i$ to vary over the sites (keeping left-right symmetry), the BI can be violated to a much higher degree. For instance, for $h_1=h_2=h_6=h_8$= 1.9, $h_3=h_4=h_5=h_a=h_b$= 0.4, $J_{ij}$ = 2.0 (these are realistic values [40]), $X_{BI}$ = 2.883, which can be compared to $2\sqrt{2} \approx 2.83$, the value for the singlet state.

If the BI is violated in a probabilistic HV system, necessarily at least one of the conditions MI, OI, and PI must be violated. Since the factorability (11) holds, OI and PI hold (as can also be calculated independently); therefore the only remaining 'resource' for violation of the BI is measurement dependence. By applying again (15), one indeed immediately finds that

$$P(\sigma_\lambda|\sigma_a,\sigma_b) \neq P(\sigma_\lambda|\sigma_{a'},\sigma_{b'}) \quad \text{for any } \sigma_\lambda \text{ and any } (\sigma_a,\sigma_b) \neq (\sigma_{a'},\sigma_{b'}) \tag{16}$$

where $\sigma_\lambda \equiv (\sigma_3,\sigma_4,...\sigma_8)$ or any subset thereof ($\sigma_i = \pm 1$). So MI in Eq. (3) does not hold, as is not really a surprise in this highly correlated system satisfying Eq. (14).

Thus this system indeed offers an example of measurement dependence *without superdeterminism* (without common causes between a, b, $\lambda$); so, if one prefers, compatible with a usual conception of free will. If this is not clear already, it can be proven as follows. In the above Gedankenexperiment Alice and Bob extract the 4 correlation functions M(a,b) needed to determine $X_{BI}$ from one total run. But if one explicitly assumes that they can *control* the value of $\sigma_a$ and $\sigma_b$, so set these spins to +1 or -1 according to their free choice, and that they perform 4 *consecutive* experiments to determine the M(a,b), then it is clear that these 4 correlation functions are identical to the ones of the first experiment (for who is not convinced, both cases can of course be explicitly calculated). Ergo, also in this second experiment the BI will be violated. In sum, this proves that correlation of λ with (a,b) through local interaction (see (16)) can arise also without assuming superdeterminism. This is important, since as we stated in Section 2, according to the accepted view [14-15, 6] violation of MI is dismissed by observing that it would be a superdeterministic fact, violating free will. This result shows once more that the conditions of probabilistic independence (MI, OI, PI) that are the premises of the BI, should be considered with extreme caution. Of course our model system seems to best apply to a Bell experiment with static analyzers; one may still wonder whether in an experiment with varying analyzer settings MI could be violated. But as we argued in the beginning of this Section, we conjecture that the existing experiments [13, 25] do not exclude that MI is violated: a resonance phenomenon may create measurement dependence through the interaction of 'something' (λ, say a field) with both analyzers.

In sum, according to this solution (S4a), local correlations at the moment of measurement are a remnant of past correlations persisting in time. Correlations that persist in time happen all the time. For briefness one could term this position '(local) measurement dependence through past interactions', a variant of what we termed supercorrelation.

**5.2. Outcome dependence (S4b)**.

Let us continue to focus on the stochastic variant of Bell's theorem. Then, as always starting from the premises (C3-C6), the next potential solution would be based on refuting outcome independence (C5); an option that would be interesting if it is compatible with Einstein locality. Note that also parameter independence (C6) could be questioned; but it is generally believed that rejecting PI amounts to a strong nonlocal effect ('non-local signaling') [4-6]. Let us now argue that OI could be violated in local physical systems. First, locality is usually supposed to imply the factorability condition (11) and OI; but recall that it has never been proven that Bell's and Einstein's locality [1] necessarily imply OI. Are there systems that are local in Bell's sense [1] (see footnote 1) but violate OI ? Closer inspection of the definitions of MI, OI and PI in (3), (8) and (9) shows that there is a relevant difference between MI and PI on the one hand, and OI on the other. As we saw in detail above, there is some justification in accepting MI and PI in experiments with spacelike separation (but see our counterarguments above). Both MI and PI involve conditional probabilities in which one of the conditioning parameters (a or b) is supposed to be irrelevant because it is chosen in a space-like separated part of the experiment. However, OI invokes a very different kind of conditional independence: here it is $\sigma_1$ that is supposed to be irrelevant for $\sigma_2$ or v.v. (given λ,a,b). But then the argument justifying MI and PI is not helpful for OI: correlation between $\sigma_1$ and $\sigma_2$ may very well be independent of the spacelike separation between the choices of a and b. Next, in the case of the singlet state of the Bell experiment, correlation between σ₁ and σ₂ (given λ,a,b) is to be expected *a fortiori*, since σ₁ and σ₂ are governed by a conservation law. Such correlation may well exist and persist over time; and the time of choice of a,b seems to have nothing to do with it. In sum, it is somewhat mysterious why OI is so generally believed to necessarily hold in local systems. Maybe because one believes that such correlations would get wiped out; but we don't know much about sub-quantum reality; and countless correlations in nature do persist almost indefinitely.

Needless to say, there is still a long way to go from this remark to the construction of a full-blown local HVT that reproduces the quantum correlations (and e.g. violates OI). However it seems again possible to corroborate above arguments by a Bell-type Gedankenexperiment on a known physical system. Indeed, consider again a spin lattice as in Fig. 1 (containing N≥ 10 lattice points), but this time described by the 'quantum Ising' Hamiltonian [43]:

$$\mathbf{H} = -J \sum_{ij} \hat{\sigma}_{z,i}\hat{\sigma}_{z,j} - h.J \sum_{i} \hat{\sigma}_{x,i}. \qquad (17)$$

This Hamiltonian describes other magnetic materials than those described by (12) [43]. Here the $\hat{\sigma}_z$ and $\hat{\sigma}_x$ are Pauli spin matrices; the first sum runs over all nearest neighbour pairs; and h is a dimensionless coupling constant. As we will show elsewhere [42], performing on this system the Bell-type correlation experiment described in Section 5.1, one finds that the BI can be violated under very broad conditions. (The HV are here the eigenvalues of $\hat{\sigma}_z$ on the sites other than 1,2,a,b [42].) One also finds that *none* of the conditions OI, MI and PI holds here, even if this system is local in Bell's original sense [1], as any known physical system is (the interactions J are truncated after first neighbours and there is obviously no superluminal interaction between parts). Note that the system does not even have a symmetry between σ₁ and σ₂; in which case violation of OI is expected a fortiori.

Based on above arguments, we conjecture that another solution to Bell's theorem exists, namely 'local outcome dependence' (S4b), which we classified as a form of supercorrelation. Realistic local HVTs should exhibit a dynamical mechanism, e.g. mediated by a field (possibly along the lines of [33-35]), that creates such a supercorrelation and reproduces the quantum correlations.

### 5.3. Supercorrelation as an intermediate position between Bohr's and Einstein's.

Let us end this Section with a brief remark, linking in a way the four positions S1-S4. From large parts of what we said above, it is clear that both superdeterminism and supercorrelation point to a causal 'connectedness' between particles (objects) and their (space-time) environment. Superdeterminism includes 'everything' in the environment, supercorrelation only typical experimental arrangements. (That is why the latter seems the more physical solution.) As we already noted in Sections 3 and 5.1, this reminds us of the well-known 'contextuality' or 'holism' of the Copenhagen interpretation and of Bohm's theory; so there is a relevant link between the four positions. Interestingly, this connection between S1-S4 seems corroborated by what Einstein [44] maintained about the EPR paradox, long after its publication. Here is what the great man said (quoted by Bell in [2]):

"If one asks what, irrespective of quantum mechanics, is characteristic of the world of ideas of physics, one is first of all struck by the following: the concepts of physics relate to a real outside world. […] It is further characteristic of these physical objects that they are thought of as arranged in a space-time continuum. An essential aspect of this arrangement of things in physics is that they lay claim, at a certain time, to an existence independent of one another, provided these objects are situated in different parts of space. The following idea characterizes the relative independence of objects far apart in space (A and B): external influence on A has no direct influence on B. […]"

"There seems to me no doubt that those physicists who regard the descriptive methods of quantum mechanics as definitive in principle would react to this line of thought in the following way: they would drop the requirement […] for the independent existence of the physical reality present in different parts of space; they would be justified in pointing out that the quantum theory nowhere makes explicit use of this requirement. I admit this, but would point out: when I consider the physical phenomena known to me, and especially those which are being so successfully encompassed by quantum mechanics, I still cannot find any fact anywhere which would make it appear likely that [that] requirement will have to be abandoned. I am therefore inclined to believe that the description of quantum mechanics […] has to be regarded as an incomplete and indirect description of reality, to be replaced at some later date by a more complete and direct one."

In a sense, then, supercorrelation is a position intermediate between Einstein's and Bohr's (or at least the position that Einstein attributes to faithful fans of the Copenhagen interpretation): it assumes that there indeed may be an interdependence between physical realities in far apart places; at the same time it aims at explaining this interdependence in a 'realist's' way à la Einstein.

In sum, the four interpretations S1-S4 have a striking element in common (they all deal with the connectedness of things); but they favor very different physical theories and very different philosophies to explain this connectedness.

## 6. Conclusion.

We reviewed here some of the admissible interpretations of Bell's theorem, paying attention to some lesser-known aspects. The most precise starting point to look for solutions to Bell's theorem are the premises (C1-C3) and (C3-C6) which are the minimal assumptions to derive the BI, in respectively a deterministic and stochastic setting [4-6]. Besides the orthodox interpretation (indeterminism) and a well-known non-standard solution (nonlocality, i.e. the existence of nonlocal influences in Bell's strong sense), we investigated two rather neglected solutions, termed here 'superdeterminism' and 'supercorrelation'. Superdeterminism rejects MI 'through common causes between λ and (a, b)'; supercorrelation rejects OI or MI 'through past interaction'. *All* these solutions have physical and metaphysical components; they *all* have something mind-boggling about them. Superdeterminism is often considered implausible by the community of quantum physicists and philosophers [2, 14-15, 19], but strictly on the basis of extra-physical arguments linked to 'free will' or conspiracy. These criticisms can be questioned simply because they are metaphysical. It was argued that superdeterminism is a solution that does not violate any known physical law, that is based on the simplest ontology, and that has therefore a strong philosophical appeal. It was also shown that supercorrelation has the additional advantage that it is more easily backed-up by physical arguments.

Indeed, we emphasized the importance of investigating the conditions MI, OI and PI in realistic physical systems, since it appears extremely difficult to assess their validity by logic or mathematics alone. It was shown that in a correlation experiment on certain spin lattices the Bell inequality can be strongly violated; yet these systems are local according to usual definitions [1,12]. This surprise was explained by the fact that these systems violate MI. They do so even if one assumes free will; one almost always supposes that MI must hold because of free will [14-15]. Consequently we argued that violation of MI and OI through local interaction cannot be excluded on the basis of existing Bell experiments, not even those with a spacelike separation between the left and right measurement events [13, 25]. Experiments with effectively changing (rotating) analyzer directions might impose MI, and would therefore be of interest. Because of the symmetry that links $\sigma_1$ and $\sigma_2$ in a singlet state, OI seems to be even less compelling than MI.

In conclusion, the Ising lattices, the early HVTs for double-slit interference we mentioned [33-35], and the remarkable experiments by Couder et al. [36] all have something in common, namely a strong correlation between all system variables, presumably even a full pairwise correlation as in the Ising lattices (see Eq. (14)). Since in such strongly correlated systems the BI may be violated, we hope that other such systems will be investigated, in particular those with a temporal dynamics and with a symmetry between the two particles.

The ultimate goal of this program would be to devise a realistic HVT for the singlet state, and beyond. Until the day such a genuine HVT explaining quantum mechanics - and predicting something new - will be confirmed by experiments, the different interpretations of Bell's theorem are likely to remain with us. In the meanwhile it seems best to take indeterminism (S1) not for more than what it is, namely one of several possible hypotheses – not an axiom of a physics theory. Other interpretations just pursue a usual scientific program, namely to explain striking probabilistic behavior, here the EPR-Bell correlations.

Acknowledgements. I would like to thank, for many stimulating discussions, Gilles Brassard, Mario Bunge, Emmanuel M. Dissakè, Henry E. Fischer, Yvon Gauthier, Gerhard Grössing, Marian Kupczynski, Jean-Pierre Marquis and Eduardo Nahmad.

**Appendix. Determinism in the history of philosophy; Spinoza's system**.

In particular concerning the issue of determinism discussed in Section 4, there is a strong interrelation between physics and philosophy. Rejecting (total) determinism (S3) amounts, in a sense, to a dramatic discontinuity in the history of western thought (which is of course not a proof of determinism). The following is a highly condensed and selective introduction to the history of this position, an introduction inevitably biased by personal preference.

Actually, consulting general encyclopedia of philosophy would suffice to see that determinism and free will belong to the most hotly debated topics of philosophy. Virtually all well-known philosophers – and countless scholars from other fields - have written about the topic [21, 45]. The debate is millennia old: among the first known western philosophers who defended determinism were Leucippus and his pupil Democritus (5$^{th}$ cent. BC), who stated that *everything happens out of necessity, not chance*. Democritus, often called the 'Father of Science', is most famous for having elaborated a detailed and incredibly modern-looking atomic theory. It is, in this context, a fascinating question to inquire on what basis Democritus could conjecture so precociously the existence of ultimate and indivisible constituents of matter, governed by laws. According to Sextus Empiricus he did so on the basis of empirical observations, such as the fact that certain substances dissolve in water in constant ratios, that certain physical and biological components degrade but also regenerate, etc.; but also of theoretical principles - namely the principle of determinism (everything has a cause) and the related idea of 'nihil ex nihilo' (nothing comes from nothing). The latter idea has been retraced to Parmenides (6$^{th}$ cent. BC), but might be much older, since it was generally accepted in Greek antiquity. It is not exaggerated to say, we believe, that these ideas are among the very few founding postulates of science and philosophy.

Since the ancient Greeks, some of the philosophers who defended determinism (understood: total determinism) were Avicenna, Spinoza, Leibniz, Locke, Kant, Schopenhauer, Laplace, Russell, Einstein, S. Hawking - to name a few. Needless to say, the views of these philosophers on determinism may differ in certain respects; but the key ingredient clearly remains. As already mentioned, the majority of philosophers who believed in determinism also believed in the absence of free will in the usual sense (but not all of them, see [45]). Aristotle was one of the early advocates of indeterminism (the action of irreducible chance or randomness). Leibniz' name is forever linked to his celebrated 'principle of sufficient reason', stipulating that everything must have a reason. Kant famously elected the thesis that all events have a cause one of his 'synthetic a priori principles'.

In the above list we believe Baruch Spinoza (1632 – 1677) deserves a special mention. We find Spinoza's defense of determinism particularly attractive and powerful, since he puts determinism at the very basis of a systematic theory of the world and of human action [30-31]. (Needless to say, strong philosophical theses can in general not be proven, but they can acquire cogency if they are part of full-blown theories that explain things; obviously, the more the theory explains, the more convincing the founding premises are.) In Spinoza's principal work, the Ethics, constructed as a deductive system based on axioms, determinism is omnipresent from the start [30-31]. Moreover, it is simple and radical. One typical example is Spinoza's Proposition 29 of the Ethics, Part 1: "Nothing is fortuitous in Nature; everything is determined by the necessity of Nature to exist and produce effects in a given manner." In other places Spinoza illustrates this thesis by stating that every individual action of any human being is as determined, as necessary to happen, as it is necessary that the sum of the angles of a triangle is 180°. In sum, within Spinoza's philosophy human free will is an illusion, or rather, should be redefined (which however does not bring Spinoza to fatalism, but to a wonderful pro-active ethical theory).

Even this very selective review will illustrate that 'measurement independence', a necessary assumption of all Bell theorems, cannot be considered as 'obvious', as is so often done. In many well-known, popular, and solid ontologies, such as Spinoza's, it does not hold. As already said, in one sense the indeterminism of S1 represents a formidable discontinuity in the history of science. S1 implies that a property $\sigma$ acquires during measurement a certain value (+1 or -1) based on no 'reason' (cause) whatsoever; and that no theory ever will be able to provide such a reason (i.e. new physical parameters of which $\sigma$ will appear to be a function, or that determine the probability $P(\sigma)$). But the history of scientific discovery is the history of finding explanations of phenomena that are only random at first sight. S1 says: we can stop our search for explanations here, forever. And yet, S1 may be the right interpretation. A good dose of agnosticism seems in place.

As a last remark, notice that for a true determinist also supercorrelation (S4) can be considered compatible with the principle of determinism, the hypothesis that everything has a cause. Indeed, suppose that a theory would exist agreeing with supercorrelation. If that theory would predict some probabilities then these may of course be supposed to result from hidden causes not part of the theory. Probability theory does not prohibit such an assumption. Indeed, one of its fathers, Laplace, believed that any probability is only a tool we need because of our ignorance of hidden causes. And the examples in physics in which probabilistic behavior can very well be retraced to deterministic laws, are countless.

Thus full determinism (S3) remains possible *at least as a philosophy*. S4 is (more easily) subject to scrutiny as a part of a physical theory. And indeed, some will find that S3 or S4 explains the 'connectedness' of things in a less mysterious way than S1 (or the Copenhagen interpretation), which essentially just accepts it.

From the point of view of philosophy, one further remarkable point is that S3 offers a solid scientific basis for theories such as Spinoza's; a possible link with oriental philosophies will not have escaped from the attention of experts. We refer to Spinoza's work [30] to remind the reader that there a subtle view on free will and human interaction is exposed. On a very personal note, it may therefore be utterly relevant for the formidable problems western society faces: the latter is based on a belief in the virtually unrestricted freedom of the individual. Which may too simple a picture.